\documentclass[%
aps,
prl,
twocolumn, 
 amsmath,amssymb, 
]{revtex4-1}

\usepackage{graphicx}
\usepackage{dcolumn}
\usepackage{bm}
\usepackage{graphicx,color}
\usepackage{dcolumn}
\usepackage{bm}
\usepackage{epstopdf}
\usepackage{verbatim}
\usepackage{enumerate}
\usepackage{textcomp}
\usepackage{filemod}

\usepackage{lineno}

\usepackage[official]{eurosym}
\usepackage[squaren]{SIunits}
\usepackage{xcolor}
\usepackage{hyperref}
\usepackage{amsmath}

\usepackage[normalem]{ulem} 

\newcommand{\Udz}{U_{\mathrm{d}}(z)}
\newcommand{\kd}{\kappa_{\mathrm{d}}}
\newcommand{\Qo}{Q_{1}(z)}

\newcommand{\stot}{\sigma_{\mathrm{t}}}
\newcommand{\str}{\sigma_{\mathrm{tr}}}
\newcommand{\sss}{\sigma_{\mathrm{s}}}
\newcommand{\sa}{\sigma_{\mathrm{a}}}

\newcommand{\ls}{l_{\mathrm{scat}}}
\newcommand{\lt}{l_{\mathrm{tr}}}
\newcommand{\ltd}{l_{\mathrm{tr}}^\mathrm{d}}
\newcommand{\la}{l_{\mathrm{abs}}}

\newcommand{\La}{L_\mathrm{a}}
\newcommand{\Lad}{L_\mathrm{a}^\mathrm{d}}

\newcommand{\FIGesmall}[5]{
	\begin{figure}[h!]
		\centering
		\includegraphics[width=#3]{#2.eps}
		\caption[#4]{\textbf{#4} #5}
		\label{#1}
	\end{figure}
}

\newcommand{\FIGesmal}[5]{
\begin{figure}[h!]
\begin{figure}[t]
		\centering
		\includegraphics[width=#3]{#2.pdf}
		\caption[#4]{\textbf{#4} #5}
		\label{#1}
	\end{figure}
}

\begin{document}


\title{Universal Validity Ranges of Diffusion Theory \\for Light and Other Electromagnetic Waves}

\author{Maryna L. Meretska}%
\author{Ravitej Uppu}%
\author{Ad Lagendijk}
\author{Willem L. Vos}
\email{w.l.vos@utwente.nl}
\affiliation{%
 Complex Photonic Systems (COPS), MESA+ Institute for Nanotechnology,
University of Twente, P.O. Box 217, 7500 AE Enschede, The Netherlands
}%

\begin{abstract}
The well-known diffusion theory describes propagation of 
light and electromagnetic waves in complex media.  
While diffusion theory is known to fail both for predominant forward scattering or strong absorption, its precise range of validity has never been established. 
Therefore we present  precise, universal limits on the scattering properties, beyond which diffusion theory yields unphysical negative energy density and negative incident flux.
When applying diffusion theory to samples outside validity ranges to {\it infer} scattering properties from transmission and reflection, the resulting transport parameters deviate by up to an order of magnitude compared to the true ones. These discrepancies are relevant to atmospheric and climate sciences, biophysics and health sciences, white LEDs and lighting, and Anderson localization of waves.
\end{abstract}

\maketitle

Understanding the propagation of light and other electromagnetic waves in scattering media with absorption and anisotropy is crucial in many areas of research~\cite{Ishimaru78, Durian91, Wiersma13,Leung14}, ranging from astrophysics to clouds~\cite{Takano89} and climate science~\cite{Fournier94, Fell01, Stramski04, Taylor05} to biology~\cite{Star88, Star89a, Cheong90,Pickering93,Dickey98, Dickey01, Klose06}, to pharmaceuticals and to sustainable energy generation~\cite{Funk73, Zege93, Sekulic96, Burger97,  Shinde99,Shi10}. 
Transport theory is used to describe the propagation of waves in these complex materials~\cite{Ishimaru78}. 
Given the complexity of transport theory, analytic solutions only exist for very simple geometries. 
In
realistic situations one usually relies on numerical approximations or on simulation techniques. 
In order to gain physical insight into the solutions of transport theory one needs to find reliable analytical approximations. 
The first-order analytic approximation to transport theory is the well-known diffusion theory~\cite{Ishimaru78}.

Diffusion theory is known to be a useful approximation to transport theory for media with a wide range of transport parameters~\cite{Takano89, Cheong90, Durian96}. 
The accepted ranges of validity are typically qualitatively described by comparing the sample thickness $d$ to several characteristic length scales~\cite{Kaveh91}:
\begin{equation} 
\lt\ll d\ll \la
 \mathrm{,}
\label{eq:fund}
\end{equation}
where the transport mean free path $\lt$ is the average distance light travels before losing information about its initial direction due to scattering, and the absorption mean free path $\la$ is the average distance light travels before being absorbed~\cite{Lagendijk96, Wiersma13, Elaloufi02, Rotter17}. 
Whereas the inequalities~(\ref{eq:fund}) provide guidelines for the range of validity, they only offer a qualitative picture. 
Therefore, the central questions we address in this paper are: 
``Are there sharply defined ranges of validity of diffusion theory?", and if so ``What happens outside such a range?". 
In brief, the answers provided by this Letter are respectively: (a) yes, there are precise ranges of validity, and (b) outside these ranges we find unphysical behavior.

Both transport theory and diffusion theory of propagation of electromagnetic waves use the characteristic scattering length scales $(\ls,\la,\lt)$\cite{Kokhanovsky15}. 
Here, the scattering mean free path $\ls$ is the average distance light travels between two consecutive collisions. 
These characteristic length scales have straightforward relations to the single-particle properties $(\sss, \sa, \mu)$~\cite{Ishimaru78}: 
\begin{equation} 
\ls=\frac{1}{\rho \sss}\equiv\frac{1}{\mu_s}
 \mathrm{,}
\label{eq:pt1}
\end{equation}

\begin{equation} 
\la=\frac{1}{\rho \sa}\equiv\frac{1}{\mu_a}
 \mathrm{,}
\label{eq:pt2}
\end{equation}

\begin{equation} 
\frac{1}{\lt}=\frac{(1-\mu)}{\ls}+\frac{1}{\la}\equiv\frac{1}{\mu_s'}+\frac{1}{\mu_a}
 \mathrm{,}
\label{eq:pt3}
\end{equation}
where $\sss$ is the scattering cross section of the scattering particles, $\sa$ is their absorption cross section. The anisotropy factor $\mu$  is defined as the average scattering angle, $\mu \equiv \langle \cos \theta \rangle$, of a scattering particle and $\rho$ is the density of scattering particles. 
The coefficients $\mu_s$, $\mu_a$, $\mu_s'$ are invoked as they are often used in the vast field of biophotonic research~\cite{vanGemert85,Keijzer89,Cheong90}.

\FIGesmall{fig:s}{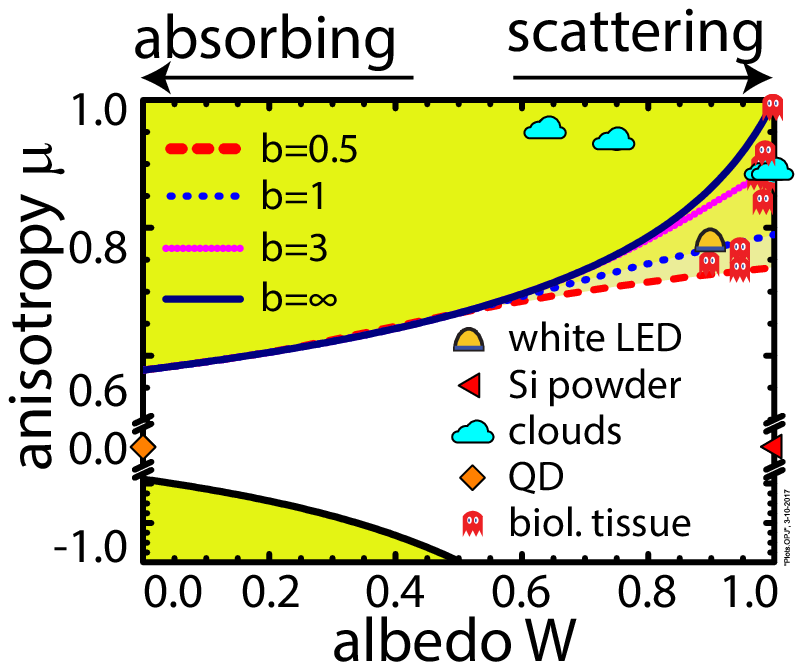}{3.5 in}{Validity range of diffusion theory for a slab shown as a 2D map.}{Lime-green regions are the areas where the diffusion equation gives unphysical results. 
The dashed and dotted curves are validity boundaries for several optical thicknesses $b$. 
The solid lines indicate the boundary for a semi-infinite slab $(b=\infty)$. 
The  symbols represent sample parameters for popular scattering media as indicated in the legend (QD refers to quantum dots)~\cite{Takano89, Cheong90, Gomez99, Leatherdale02, Meretska17}.}

The main advantage of diffusion theory is that simple and analytical solutions are offered for ubiquitous geometries such as the finite slab and the sphere. 
For instance, for a plane wave incident upon a slab with thickness $d$ the diffusion equation for the diffuse energy density $\Udz$ has the form~\cite{Ishimaru78}
\begin{equation} 
\frac{\partial^2}{\partial z^2} \Udz-\kd^2 \Udz=- Q_{0} \exp(-\rho  \stot z)
 \mathrm{,}
\label{eq:difeq}
\end{equation}
where $Q_{0}\equiv\lbrack 3 \rho \sss \rho \str+3 \rho \sss \rho \stot \mu \rbrack(F_0 /4\pi)$, $F_0$ is the magnitude of the incident flux, $\str =(1-\mu)\sss+\sa$ is the transport cross section,  $\kd^2 \equiv 3\sa\str$, and $\stot \equiv \sss+\sa$ is the total cross section. The mixed boundary conditions are
\begin{eqnarray}
\Udz-h\frac{\partial}{\partial z}\Udz +\frac{\Qo}{2\pi}=0 & \text{ }&\text{        at $z=0$} 
 \mathrm{,}
\label{eq:bound1}\\
\Udz+h\frac{\partial}{\partial z}\Udz -\frac{\Qo}{2\pi}=0 & \text{ }&\text{        at $z=d$}
 \mathrm{,}
\label{eq:bound2}
\end{eqnarray}
where $\Qo\equiv\sss \mu \str F_0 \exp(-\rho \stot z)$ and $h\equiv 2/\left(3\rho\str\right)$. 
The solution to differential equation~(\ref{eq:difeq}) with boundary conditions~(\ref{eq:bound1}) and (\ref{eq:bound2}) is the starting point of our discussion. We have verified the solution given in the literature\cite{Ishimaru78},  both analytically and numerically.
The diffuse energy density  $\Udz$ is a positive definite scalar, a condition that is rigorously expressed as
\begin{eqnarray}
\Udz\ge 0 & \text{ }&\text{        $\forall$ $z$ $\in$ $\left[ 0,d\right]$} \label{eq:ubound} 
 \mathrm{.}
\end{eqnarray}
The diffuse flux ${\bm F}_{\rm d}(z)\equiv F_{\rm d}(z) {\bm {\hat z}}$ is a one-dimensional vector~\cite{Ishimaru78}, whose outgoing direction changes from leftward, the incoming direction, at the front surface to rightward at the back surface of the slab, conditions that are expressed rigorously as
\begin{eqnarray}
F_{\rm d}(z)<0 & \text{ }&\text{        at $z=0$} \label{eq:fboundleft}
 \mathrm{,}
\\
F_{\rm d}(z)>0 & \text{ }&\text{        at $z=d$}
 \mathrm{.}
\label{eq:fboundright}
\end{eqnarray}
The rigorous conditions (\ref{eq:ubound}-\ref{eq:fboundright}) replace the qualitative inequalities~(\ref{eq:fund}).
When any of these conditions (\ref{eq:ubound}-\ref{eq:fboundright}) is violated the diffusion equation fails, as observable quantities then display unphysical behavior, namely negative reflection or transmission coefficients or a negative energy density.

To check the validity range of the diffusion equation for the slab and the semi-infinite slab we explored the full parameter space. 
While several independent parameter sets can fully characterize a sample, we choose the set consisting of the three dimensionless parameters ($W, \mu, b$), where the albedo $W$ is defined as $W \equiv \sss/(\sss+\sa)$,  the dimensionless optical thickness of the slab $b$ is defined as $b\equiv d/\ls$, and the anisotropy factor $\mu$ was defined above.

Our main results are summarized in Fig.~\ref{fig:s}. This phase diagram shows two lime-green regions that indicate the part of the parameter space where diffusion theory fails, as it generates a negative energy density. 
In the white region the results of  diffusion theory are mathematically sound.    
It appears that the boundary conditions on the flux $F(z)$ and the energy density $U(z)$ are equivalent:
in Fig.~\ref{fig:s} the two boundaries of the applicability range of the diffusion theory were calculated from the vanishing of the energy density:
\begin{eqnarray}
U_\mathrm{d}(0) &=& 0,  \text{ }\text{at the right side of the slab $z=0$}\mathrm{,}
\label{eq:ineq0}\\
U_\mathrm{d}(d) &=& 0,  \text{ }\text{at the left side of the slab $z=d$}\mathrm{.}
\label{eq:ineqd}
\end{eqnarray}
The equivalent conditions for the validity of the flux are
\begin{eqnarray}
F(0) = 0, & \text{ }&\text{at the right side of the slab $z=0$}\mathrm{,}
\label{eq:ineqtF}\\
F(d) = 0, & \text{ }&\text{at the left side of the slab $z=d$}\mathrm{.}
\label{eq:ineqbF}
\end{eqnarray}

Naively, one may expect the boundaries defining the validity of the diffusion equation to depend on the optical density. 
We find, however, that only the top unphysical region depends significantly on the optical thickness as shown in Fig.~\ref{fig:s}. 
For an albedo $W>0.5$, the top unphysical region is more significant for greater optical thickness $b$. 
When $W<0.5$, we observe no changes in the top unphysical area for varying optical thickness $b$. 
This insensitivity for strongly absorbing samples is explained by the fact that light does not reach the right side of the slab. In Figure~\ref{fig:s} the locations of some widely-studied scattering media are indicated. 

\FIGesmall{fig:s11}{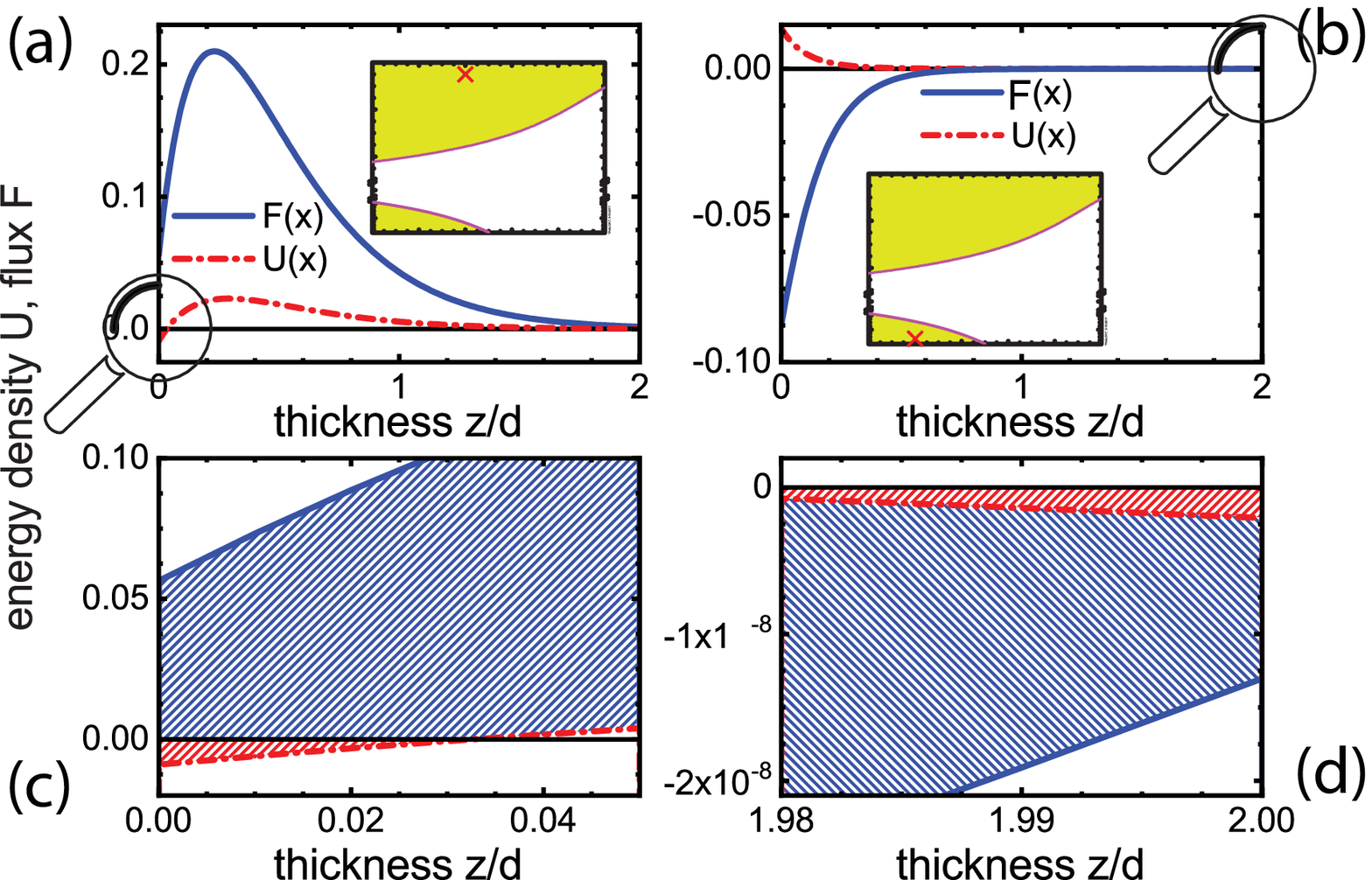}{3.5 in}{Energy density (red curves) and flux (blue curves) in the unphysical range of transport parameters.}
{\textbf{(a)} In a strongly forward scattering and absorbing material  the unphysical conditions occurs on the left boundary: an inward flux $F(0) > 0$ and a negative energy density $U(0) < 0$. 
\textbf{(b)} In a strongly backward scattering material with strong absorption  the unphysical conditions occur at the right boundary of the slab: an inward flux $F(d) < 0$ and a negative energy density $U(d) < 0$. 
Both insets show the conditions in the parameter space as red crosses, with  $b=3$. 
\textbf{(c,d)} Magnified parts of  Figs.~\ref{fig:s11}(a,b), respectively, as  indicated with magnifying glasses in Figs.~\ref{fig:s11}(a,b). Hatched areas refer to unphysical values. 
The flux is scaled with $F_0$, the magnitude of the incident flux and the energy density is scaled with $F_0/4\pi$.} 

To illustrate the wrong predictions when the diffusion theory is applied to samples with parameters in the unphysical range, we show in Fig.~\ref{fig:s11} the energy density and flux for two such samples. 
For a strongly forward scattering and absorbing material, Figure~\ref{fig:s11}(a) shows that the flux and energy density are unphysical near the left (entrance) boundary of the sample. 
A real-world physical situation where this behavior arises are certain types of clouds, see Fig.~\ref{fig:s}.
For a strongly backward scattering and absorbing sample, Fig.~\ref{fig:s11}(b) shows that the flux and energy density are unphysical near the right (exit) boundary of the sample. 

\FIGesmall{fig:s2}{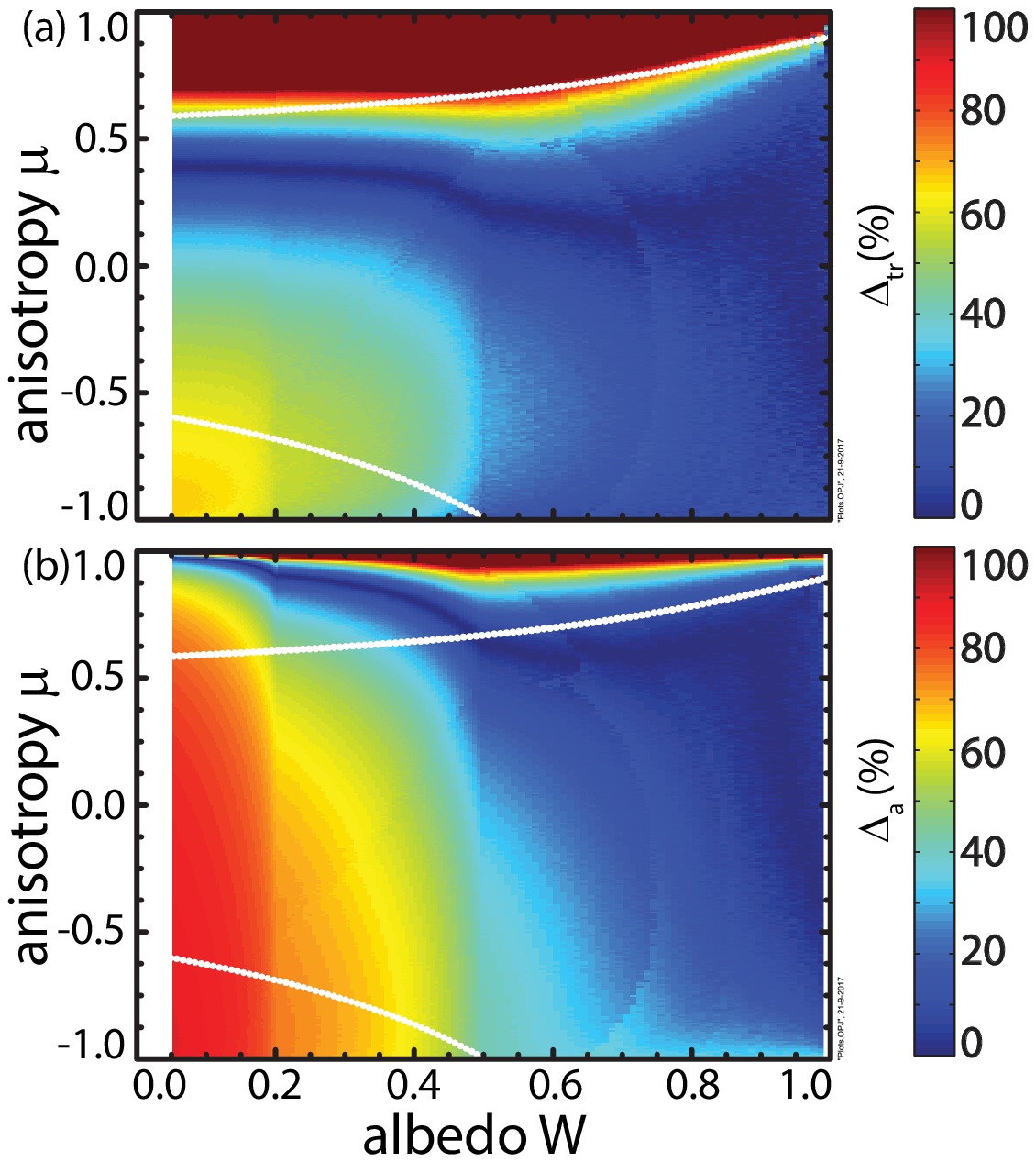}{3.5 in}{Relative error in diffusion theory calculations shown as a 2D map.}{ The white curves show the analytically derived boundary for the range of validity of the diffusion theory for optical thickness $b = 3$. 
\textbf{(a)} Color gradient indicates the relative error $\Delta_{\mathrm{tr}}$ in the apparent mean free path $\lt^{\rm d}$ obtained using the diffusion theory compared to the rigorous $\lt$. 
\textbf{(b)} Color gradient indicates the relative error $\Delta_{\mathrm{a}}$ in the apparent absorption length $\la^{\rm d}$  using the diffusion theory compared to the rigorous $\la$.
}

We now discuss how scientists who experimentally study samples with parameters in an unphysical range of diffusion theory apparently manage to interpret their results with diffusion theory without encountering any problem. 
From an experimental point of view, a scattering sample is typically characterized with the set of three length scales
$(d,\lt, L_\mathrm{a})$, where the absorption length $L_\mathrm{a}$ is defined as $L_\mathrm{a}\equiv \sqrt { \lt\la/3}$, which can be associated with the average distance between the start and the end of a random-walk with absorption mean free path $\la$. 
The set  $(d,\lt, L_\mathrm{a})$ is fully equivalent to the set ($W, \mu, b$). 
We emphasize that both sets always represent a valid set of parameters to characterize wave transport in a scattering sample, irrespective of the validity of diffusion theory. 
In a typical experiment the total transmission coefficient $T$ and the total reflection coefficient $R$ are measured. 
From these experimental data, the length scales ($\ltd$, $\Lad$) are inferred using diffusion theory, and no unphysical behavior is noticed for samples having ($W,\mu, b$) far in the unphysical regions. 
However, in such cases the \textit{extracted length scales} ($\ltd$, $\Lad$) correspond to values for ($W^{\rm d},\mu^{\rm d},b^{\rm d}$) that \textit{differ strongly from the true values} ($W,\mu, b$).

We now investigate quantitatively how much the apparent transport parameters deviate from the true ones, when using diffusion theory in regions where it is unphysical. 
 To obtain the true parameters we have performed extensive Monte Carlo simulations, using the method described in Ref.~\cite{Mujumdar2010JNP, Uppu2013PRA}.
In the simulations, samples are characterized by the parameter set ($b$, $\lt$, $L_\mathrm{a}$). 
For several thicknesses $b$ we calculate the transmission $T$ and the reflection $R$ for many values of the parameter pair $(W,\mu)$ and convert these values onto the ($\lt,\La$) parameter space. 
In this way we obtain for a specific thickness $b$ an exact mapping of the parameters ($\lt,\La$) to the observables ($T$, $R$), and \textit{vice versa}. 
Given the set of observables ($T$, $R$) obtained from our simulations, we then calculate {\it with diffusion theory} the apparent length scales $(\ltd,\Lad)$ that would produce these observables $T$ and $R$. 
These apparent length scales are then compared with the true parameters $(\lt,\La)$. 
To evaluate the relative error of the apparent length scales obtained with diffusion theory we define the relative errors
\begin{equation} 
\Delta_{\mathrm{tr}} \equiv \frac{\mid \lt -\ltd \mid}{\ltd}
 \mathrm{,}
\label{eq:errt}
\end{equation}
and
\begin{equation} 
\Delta_{\mathrm{a}} \equiv \frac{\mid \La -\Lad \mid}{\Lad}
\label{eq:erra}
\end{equation}
for the transport and the absorption mean free paths, respectively.
In Fig.~\ref{fig:s2} we show for an optical thickness $b=3$ the full 2D map of the relative errors for both the transport mean free path $\Delta_{\mathrm{tr}}$ and the absorption length $\Delta_{\mathrm{a}}$. 
In the same figure we have indicated the boundaries separating the two unphysical regions from the physical region of  diffusion theory. 
These boundaries, calculated from boundary conditions (\ref{eq:ubound}) to (\ref{eq:fboundright}),  divide the scattering parameter space into three parts: an upper unphysical region, a physical region in the center, and a lower unphysical region.

Figure~\ref{fig:s2} shows that in the upper unphysical region the relative errors in the transport mean free path are huge for all albedos, ranging from $\Delta_{\mathrm{tr}} = 100\%$ up to 1000\%. 
The corresponding relative errors in the absorption mean free path are also substantial, ranging from $\Delta_{\mathrm{tr}} = 50\%$ to 100\%. 
We note that the illustrative sample presented in Fig.~\ref{fig:s11}(a,c) also has large relative errors for the length scales, namely $\Delta_{\mathrm{tr}}=1056\%$ and $\Delta_{\mathrm{a}}=115\%$.

In the lower unphysical region in Fig.~\ref{fig:s2} the transport mean free path has relative errors from $\Delta_{\mathrm{tr}} = 60\%$ up to 80\% and the absorption mean free path has errors up to about $\Delta_{\mathrm{a}} = 90\%$. 
In the physical region the errors are, as to be expected, much less. 
For an albedo $W \ge 0.5$ the relative errors are less than 10\%, except that close to the upper boundary the relative errors increase to about $\Delta_{\mathrm{tr}} = 100\%$. 
For more strongly absorbing samples with $W < 0.5$, the errors become substantial even in the physical region: for transport  $\Delta_{\mathrm{tr}} = 60\%$ and for absorption up to $\Delta_{\mathrm{a}} = 100\%$. 

In summary, we have derived the fundamental ranges of validity of diffusion theory in media with absorption and anisotropic scattering. 
We identify two large unphysical regions in the ($W, \mu$)-plane, where the albedo $W$ characterizes the amount of absorption and where $\mu$ represents the anisotropy parameter.
We found that a major limitation when using diffusion theory to analyze experimental reflection and transmission data is that it is reliable only with \textit{a priori} knowledge that the sample has transport parameters well inside the physical region of the ($W, \mu$)-plane. 
Otherwise, diffusion theory gives unreliable results when interpreting data on such samples.
These errors are likely to have major impact on applications with important associated societal consequences.  

Figure~\ref{fig:s} shows that widely-studied scattering media such as biological tissue, clouds, or white LEDs~\cite{Takano89, Cheong90, Meretska17} have transport parameters near or even within the upper unphysical range. 
Erroneous transport parameters for clouds will have consequences for atmospheric and climate sciences, in case of biological tissue there may be consequences for health sciences, and in case of white LEDs there are consequences for the energy efficiency of modern lighting. 

Our new insights have also consequences for the long-sought Anderson localization where diffusion theory breaks down due to interference, either for light, for microwaves, or for electrons~\cite{John1984PRL, Lag09}. 
The current consensus is that absorption is an unwanted side effect that has otherwise no consequences for the breakdown of diffusion theory~\cite{Gen91}. 
Based on our results, however, it is clear that when reporting breakdown of diffusion of electromagnetic waves one must ensure that the samples have parameters well in the physical region of the ($W, \mu$)-plane. 

\begin{acknowledgments}
It is a pleasure to thank Shakeeb Bin Hasan, Pepijn Pinkse, and Wilbert IJzerman for useful discussions. 
This work was supported by the Dutch Technology Foundation STW (contract no. 11985), by the FOM program "Stirring of Light!", and by the Dutch Funding Agency NWO.
\end{acknowledgments}
\bibliography{References}

\end{document}